\documentclass[twocolumn,aps,showpacs,pra,superscriptaddress]{revtex4-1}
\usepackage{amssymb,amsmath,amsfonts,epsfig,graphicx,latexsym,bm,color,braket,subfigure}

\begin{document}

\title{Dynamical density wave order in an atom-cavity system}

\author{Christoph Georges}\altaffiliation{These authors contributed equally to this work}
\affiliation{Zentrum f\"ur Optische Quantentechnologien and Institut f\"ur Laser-Physik, 
Universit\"at Hamburg, 22761 Hamburg, Germany}

\author{Jayson G. Cosme}\altaffiliation{These authors contributed equally to this work}
\affiliation{Zentrum f\"ur Optische Quantentechnologien and Institut f\"ur Laser-Physik, 
Universit\"at Hamburg, 22761 Hamburg, Germany}
\affiliation{The Hamburg Center for Ultrafast Imaging, Luruper Chaussee 149, Hamburg 22761, Germany}
\affiliation{National Institute of Physics, University of the Philippines, Diliman, Quezon City 1101, Philippines}

\author{Hans Ke{\ss}ler}
\affiliation{Zentrum f\"ur Optische Quantentechnologien and Institut f\"ur Laser-Physik, 
Universit\"at Hamburg, 22761 Hamburg, Germany}
\affiliation{The Hamburg Center for Ultrafast Imaging, Luruper Chaussee 149, Hamburg 22761, Germany}

\author{Ludwig Mathey}
\affiliation{Zentrum f\"ur Optische Quantentechnologien and Institut f\"ur Laser-Physik, 
Universit\"at Hamburg, 22761 Hamburg, Germany}
\affiliation{The Hamburg Center for Ultrafast Imaging, Luruper Chaussee 149, Hamburg 22761, Germany}

\author{Andreas Hemmerich}
\affiliation{Zentrum f\"ur Optische Quantentechnologien and Institut f\"ur Laser-Physik, 
Universit\"at Hamburg, 22761 Hamburg, Germany}
\affiliation{The Hamburg Center for Ultrafast Imaging, Luruper Chaussee 149, Hamburg 22761, Germany}

\begin{abstract}
We theoretically and experimentally explore the emergence of a dynamical density wave order in a driven dissipative atom-cavity system. A Bose-Einstein condensate is placed inside a high finesse optical resonator and pumped sideways by an optical standing wave. The pump strength is chosen to induce a stationary superradiant checkerboard density wave order of the atoms stabilized by a strong intracavity light field. We show theoretically that, when the pump is modulated with sufficient strength at a frequency $\omega_{d}$ close to a systemic resonance frequency $\omega_{>}$, a dynamical density wave order emerges, which oscillates at the two frequencies $\omega_{>}$ and $\omega_{<} = \omega_{d} - \omega_{>}$. This order is associated with a characteristic momentum spectrum, also found in experiments in addition to remnants of the oscillatory dynamics presumably damped by on-site interaction and heating, not included in the calculations. The oscillating density grating, associated with this order, suppresses pump-induced light scattering into the cavity. Similar mechanisms might be conceivable in light-driven electronic matter.
\end{abstract}

\maketitle

The focus of research in many-body physics is presently shifting towards dynamical scenarios far from thermal equilibrium, often in presence of coupling to a bath and external driving, for example, by means of light. This typically adds considerable complexity to the already complex world of many-body systems rewarded by the possibility to discover novel unexpected physics with intriguing applications. The dynamical control of solids via optical driving, with the overarching goal of creating unique functionalities, has become an active and exciting field \cite{Tokura2017, Basov2017, Tobey2008, Patel2016, Fausti2011, Zong2018, Maschek2018, Kogar2019}. Another strain of research is devoted to time-crystals, i.e., driven dynamical many-body states, which break time-translation symmetry at a frequency merely determined by inherent system parameters, however, robust against the variation of external quantities \cite{Wilczek2012, Sacha2018, Else2019}. Understanding the fundamental mechanisms of driven many-body systems can considerably benefit from studying precisely controlled simplified model systems based upon ultracold atoms, particularly, if light-induced driving is applied \cite{Basov2017, Eisert2015}.

An ultracold gas of atoms inside a high-finesse optical standing wave cavity is a versatile, well controlled yet simple platform for exploring many-body physics in the presence of dissipation and driving \cite{Ritsch2013, Klinder2015, Landig2016}. The particular interest in this system is due to the cavity-induced infinite-range interaction among the atoms. For sufficiently strong transverse pumping, this system undergoes a phase transition from a spatially homogeneous Bose-Einstein condensate (BEC) phase into a self-organized density wave (DW) phase, where the atoms form a stationary checkerboard density pattern that scatters photons into the cavity mode akin to a Bragg grating \cite{Domokos2002, Nagy2008, Baumann2010, Klinder2015, Landig2016}. Significant attention has been also devoted to dynamical phenomena in the atom-cavity platform \cite{Keeling2010, Bhaseen2012, Piazza2015, Kessler2019, Chiacchio2019, Dogra2019, Cosme2018, Molignini2018, Chitra2015, Gong2018, Cosme2019, Zhu2019, Zupancic2019}. In a recent experiment \cite{Georges2018} it was shown, that by adding a second frequency component to the pump, detuned from all systemic resonances, the checkerboard DW phase is suppressed and the coherent BEC phase is partially restored. Theoretically, this scenario could be understood in terms of Floquet driving in the high-frequency regime \cite{Cosme2018}, and a possible analogy to the mechanism of light-induced superconductivity in Ref.~\cite{Fausti2011} has been discussed. 

\begin{figure}[!htb]
\centering
\includegraphics[width=1.0\columnwidth]{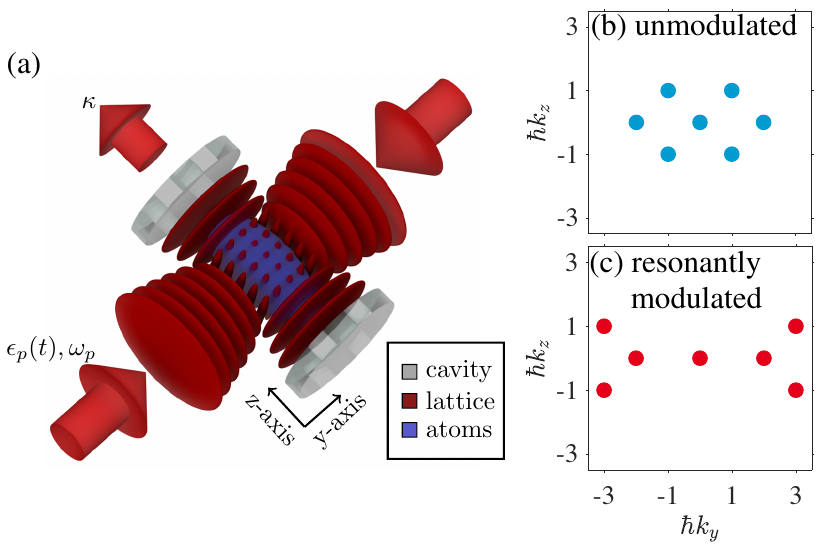}
\caption{(a) A BEC inside a high-finesse cavity is pumped along the transverse direction above a critical strength, such that a checkerboard density wave order is formed with the momentum distribution schematically illustrated in (b). If the pump is modulated near-resonantly, an oscillation between the momentum spectra in (b) and (c) arises.}
\label{fig:intro}
\end{figure}

In this work we study, theoretically and experimentally, near-resonant driving of the atom-cavity system, shown in Fig.~\ref{fig:intro}(a), initially prepared in the checkerboard DW state, which possesses a momentum spectrum schematically illustrated in Fig.~\ref{fig:intro}(b). According to calculations, a new dynamical DW state emerges, which periodically switches between two distinct higher order matter gratings (cf. Appendix B), associated with the same momentum spectrum sketched in Fig.~\ref{fig:intro}(c). Its resonance frequency $\omega_{>}$ depends on inherent system parameters as well as on the driving strength $f_{d}$ and the driving frequency $\omega_{d}$. The value of $\omega_{>}$ typically lies between $10\,\%$ and $15\,\%$ above that of the excitation frequency $\omega_{3,1} = (3^2+1^2) \,\omega_{\textrm{rec}}$ of the bare atomic momentum states $\{\hbar k_y,\hbar k_z\}=\{\pm 3, \pm 1\} \hbar k$ (with $k$ denoting the wave number of the pump beam, $m=$ atomic mass, and $\omega_{\textrm{rec}} \equiv \hbar^2 k^2/(2m)$ = recoil frequency). In contrast to the chequerboard DW, the new matter gratings suppress light scattering into the cavity due to destructive interference, which relates to the phenomenon of subradiant scattering \cite{Wolf2018}. The dynamical DW state performs an additional oscillation at the much lower frequency $\omega_{<} = \omega_{d} - \omega_{>}$. This slow oscillation is associated with a periodic change of the momentum spectrum between the two cases sketched in Fig.~\ref{fig:intro}(b) and (c). For technical reasons, only the slow oscillation at frequency $\omega_{<}$ provides a signature that we can experimentally detect. Our experimental system is strongly damped due to sizable contact interaction, parametric heating and atom loss, not included in our calculations. Nevertheless, guided by the theoretical predictions, a single oscillation cycle can be identified such that $\omega_{<}$ can be roughly determined.

\begin{figure}[!htb]
\centering
\includegraphics[width=1.0\columnwidth]{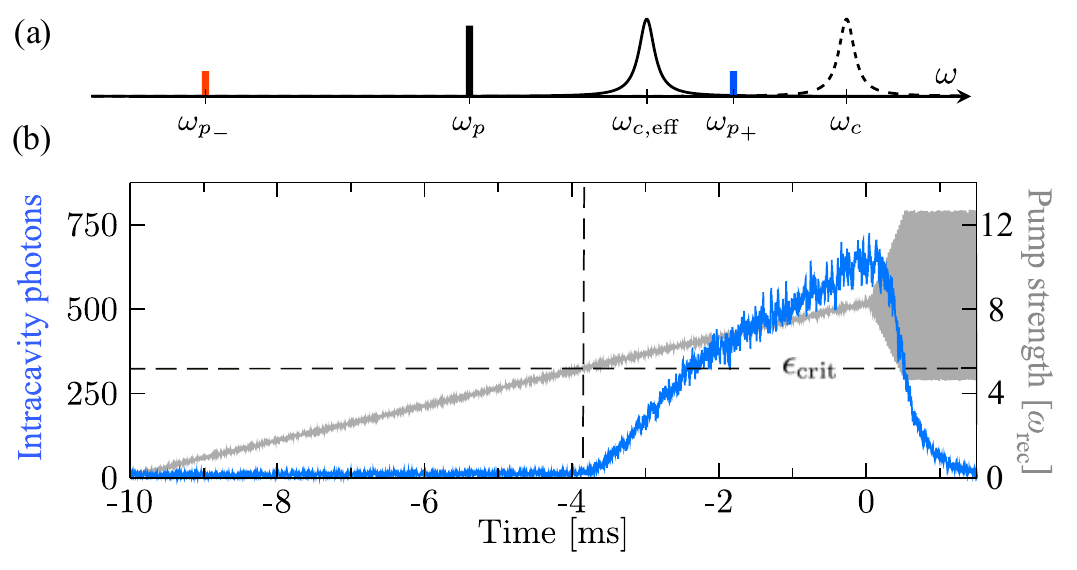}
\caption{ (a) Illustration of the relevant frequencies. (b) Experimental driving protocol (gray trace): first, the pump intensity is ramped up over 10~ms. Next, at $t=0\,$ms, the modulation strength is linearly ramped up during 0.5~ms to $f_d = 0.425$ and kept constant thereafter. The modulation frequency is $\omega_d/2\pi = 40\,$kHz.}
\label{fig:mod} 
\end{figure} 

In our experiment, a BEC of $^{87}$Rb atoms is coupled to the light field inside a high finesse cavity as sketched in Fig.~\ref{fig:intro}(a). The details are described in Ref.~\cite{Klinder2015}. First, we prepare an elongated BEC with $N_a \approx 10^5$ atoms in the $|F=2,m_f=2\rangle$ state, held in a magnetic trap  \cite{Esslinger1998}. The BEC is magnetically transported into the TEM$_{00}$ mode of the cavity, which has a beam waist of $w_0 \approx 31.2~\mu \mathrm{m}$ and a finesse of $\mathcal{F}=3.44\times 10^5$. An optical standing wave, perpendicular to the cavity axis ($z$-axis) with linear polarization perpendicular to the cavity axis, pumps the atoms. The pump wavelength $\lambda = \mathrm{803~nm}$ is far detuned from the relevant atomic transitions of $^{87}$Rb, which are the D$_1$ and D$_2$ lines at $\mathrm{795~nm}$ and $\mathrm{780~nm}$, respectively, and hence the atom-light coupling is dispersive. The pump strength $\epsilon_{p}$ is measured in terms of the recoil frequency $\omega_{\mathrm{rec}}= 2\pi \times 3.55\mathrm{~kHz}$. Due to the specifics of the $^{87}$Rb level scheme, the dominant coupling arises for left circular polarization. The TEM$_{00}$ resonance frequency for left circular polarization is dispersively shifted by an amount $\delta_{-} = \frac{1}{2}N_a U_0$, where the light shift per photon is $ U_0\approx - 2\pi \times 0.36\mathrm{~Hz}$. For $10^5$ atoms, the system operates in the regime of strong cooperative coupling, i.e., $\delta_{-} \approx -2\pi \times 18~\mathrm{~kHz}$ exceeds the cavity field decay rate $\kappa\approx 2\pi \times 4.5 \mathrm{~kHz}$. The timescales of the dynamics of the atoms and the cavity photons, $\omega_{\mathrm{rec}}^{-1}$ and $\kappa^{-1}$, respectively, are comparable. In this regime, the adjustment of the intra-cavity field to a change of the atomic distribution, and hence the cavity-mediated interaction between distant atoms, is delayed by an amount comparable with the typical time for a change of the atomic distribution. In this sense, a \textit{retarded} infinite-range interaction arises. For the pump frequency $\omega_p$, we choose a fixed value $\omega_p = \omega_{c,\textrm{eff}} - 2\pi \times 30~\mathrm{~kHz}$, where $\omega_c$ is the resonance frequency of the empty cavity and $\omega_{c,\textrm{eff}} \equiv \omega_c + \delta_{-}$ is the cavity resonance shifted by the coupling to the atoms (cf. Fig.~\ref{fig:mod}(a)). 

To prepare the system in the checkerboard phase, we linearly increase the pump strength for 10~ms and monitor the intracavity photon number. Around -4~ms, the pump strength $\epsilon_{p}$ (cf. gray trace in Fig.~\ref{fig:mod}(b)) surpasses the critical value $\epsilon_\mathrm{crit} \approx 5\,\omega_\mathrm{rec}$ and a rapid increase of the intracavity photon number (cf. blue trace in Fig.~\ref{fig:mod}(b)) indicates the transition into the checkerboard phase, where photons from the pump are scattered into the cavity. This phase is characterized by a self-organized DW pattern at wavevector $\mathbf{k} = \{k_y,k_z\}= \{1,1\} k$ and hence referred to as DW$_{1,1}$ order. This results in a substantial number of atoms $n_{\mu,\nu}$ in the momentum modes at momenta $\{\mu,\nu\} \hbar k$, with $\mu,\nu=\pm 1$. The pump strength is further increased to $\epsilon_0 \approx 8\,\omega_{\mathrm{rec}}$ to prepare the atoms well within the DW$_{1,1}$ phase. 

In the work of Ref.~\cite{Georges2018}, instead of applying modulation, we added a single sideband to the pump field to ensure that the intended dynamical suppression of the DW$_{1,1}$ phase is not due to a mere depletion of the pump. Here, since we are interested in the near-resonant excitation of higher order density waves, we may apply a simple intensity modulation protocol. Periodic modulation of the pump intensity according to $\epsilon_{p}(t) = \epsilon_0 (1 + f_{d} \,\mathrm{cos}(\omega_{d} t))$ leads to frequency sidebands $\omega_{p\pm} \equiv  \omega_{p} \pm \omega_{d}$ (cf. Fig.~\ref{fig:mod}(a)). Without inducing  excessive heating we can access modulation strengths in the range of $f_{d} \in [0,0.7]$ in the frequency range $\omega_{d}/(2\pi) \in [30,50] \mathrm{~kHz}$. The driving sequence follows the gray trace in Fig.~\ref{fig:mod}(b): at $t=0$, the modulation strength is ramped up linearly during $0.5\,$ms to the desired strength $f_{d}$ and subsequently is kept fixed for up to $0.75\,$ms. As seen in Fig.~\ref{fig:mod}(b), the modulation leads to a suppression of the intracavity photon number. Finally, after a ballistic expansion of $25\mathrm{~ms}$ duration, the momentum mode populations $n_{\mu,\nu}(f_{d})$ are extracted from an absorption image and the the relative occupations $F_{\mu,\nu}(f_{d})=n_{\mu,\nu}(f_{d})/n_{0,0}(0)$ are determined.

\begin{figure}[!htb]
\includegraphics[width=1\columnwidth]{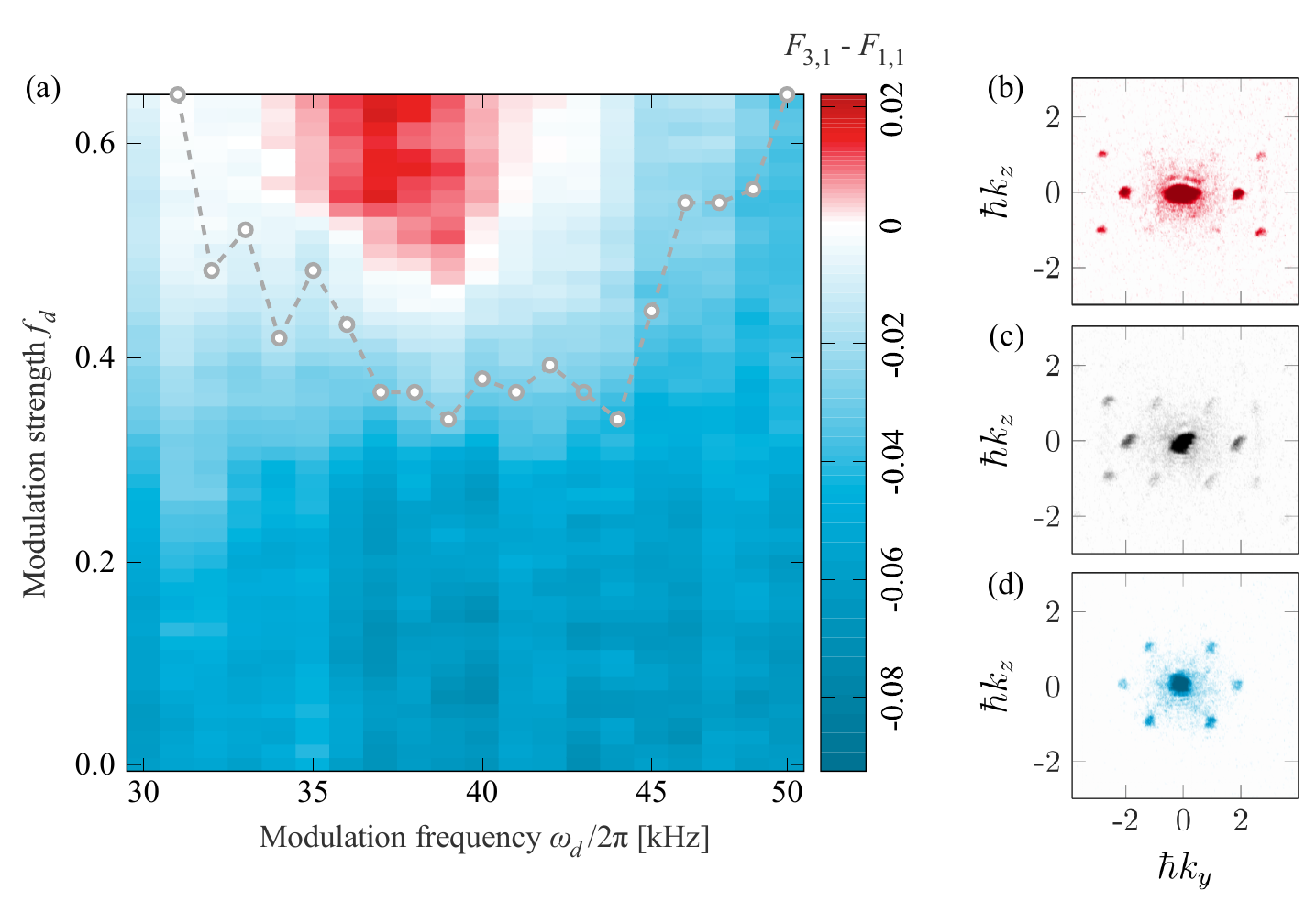}
\caption{(a) Plot of $F_{3,1} - F_{1,1}$. For weak driving ($f_{d} < 0.3$), the DW$_{1,1}$ phase is dominant (blue area). For strong driving ($f_{d} > 0.5$) within the red colored area the DW$_{3,1}$ phase prevails. The gray circles connected by dashed lines indicate the boundary above which $F_{3,1}$ exceeds $25 \%$ of its maximal value. On the right edge momentum spectra are shown for $\omega_{d}/(2\pi)= 40\mathrm{~kHz}$ with (b) $f_{d}=0.60$, (c) $f_{d}=0.47$ and (d) $f_{d}=0.10$.}
\label{fig:ExpPhDg} 
\end{figure} 

\begin{figure}[!htb]
\includegraphics[width=0.9\columnwidth]{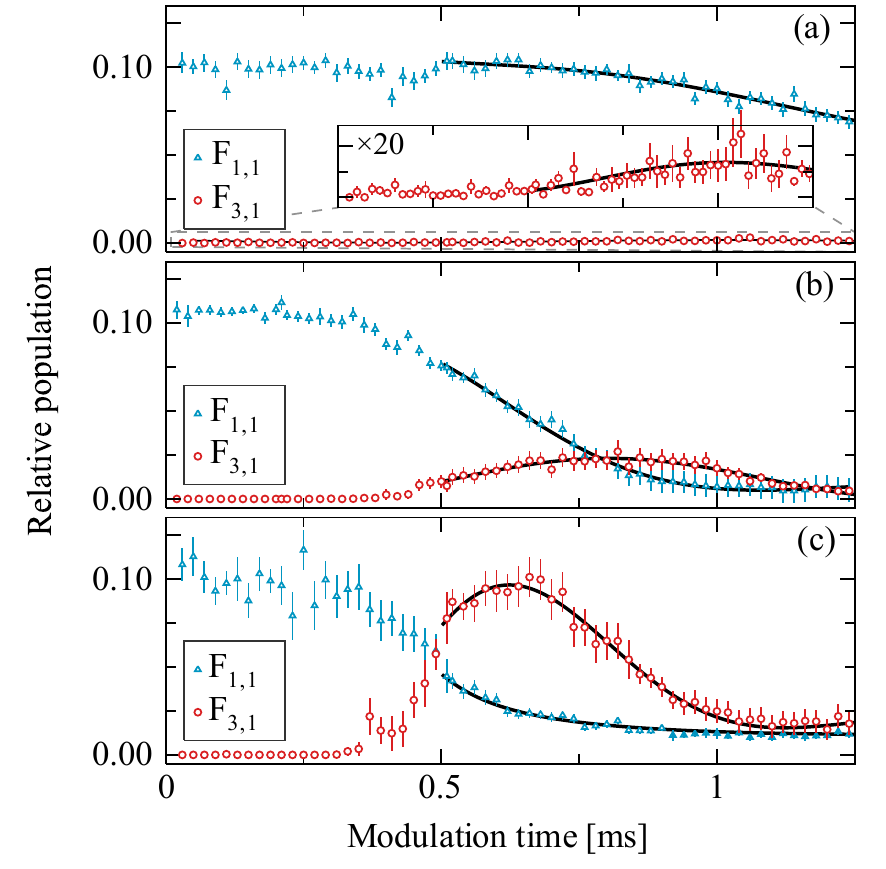}
\caption{The time evolution of $F_{1,1}$ and $F_{3,1}$ is shown for $\omega_{d}/(2\pi) = 40\,$kHz. The modulation strength is $f_{d} = 0.17$ in (a), $f_{d} = 0.52$ in (b) and $f_{d} = 0.71$ (c). The solid lines show fits with exponentially decaying harmonic oscillations with frequencies $\omega_{<} \approx 0.8\,$kHz in (a), $\omega_{<} \approx 0.9\,$kHz in (b), and $\omega_{<} \approx 1.2\,$kHz in (c).}
\label{fig:ExpPopEv} 
\end{figure} 

In Fig.~\ref{fig:ExpPhDg}(a), we map out an experimental phase diagram by plotting $F_{3,1} - F_{1,1}$, observed after linearly ramping up the modulation strength at the frequency $\omega_{d}$ during 0.5~ms to a maximum value $f_{d}$ and a subsequent waiting time of 0.5~ms. For $f_{d} < 0.3$ and outside the shown interval $\omega_{d}/(2\pi)\in [30,50]\,$kHz, upon increase of $f_{d}$, the DW$_{1,1}$ phase and the intracavity photon number are depleted, and the BEC phase is recovered, in agreement with the findings for single sideband driving in Ref.~\cite{Georges2018}. Within the red region in Fig.~\ref{fig:ExpPhDg}(a), we observe the emergence of a dynamical DW phase, indicated by positive values of $F_{3,1} - F_{1,1}$ and the momentum spectrum shown in Fig.~\ref{fig:ExpPhDg}(b). At $f_{d} < 0.3$, the DW$_{1,1}$ phase prevails, indicated by the momentum spectrum shown in Fig.~\ref{fig:ExpPhDg} (d), while in the white region the spectrum in (c) is found. Although theoretically predicted in Ref.~\cite{Cosme2018}, because of driving-induced heating for $f_{d}>0.6$, other higher order DW phases cannot be observed such that we focus here on the DW order associated with the momentum spectrum in Fig.~\ref{fig:ExpPhDg}(b) and the wavevector $\mathbf{k}=\{3,1\} k$. In contrast to the DW$_{1,1}$ phase, this order comes with a density grating that does not satisfy the Bragg condition for constructive 90 degree scattering of pump photons into the cavity. Hence, the occurrence of an intra-cavity light-field is suppressed by destructive interference, as seen in Fig.~\ref{fig:mod}(b). Nevertheless, the atoms remain under the effect of the pump standing wave potential, whose periodicity is incompatible with that of the density grating, hence giving rise to a structural instability. This drives the system into a non-stationary phase \cite{Keeling2010, Bhaseen2012, Piazza2015, Kessler2019, Chiacchio2019, Dogra2019, Cosme2018, Cosme2019} that we refer to as DW$_{3,1}$. In contrast to DW$_{1,1}$, this phase is not captured within the often applied mapping of the atom-cavity system onto the Dicke model \cite{Kirton2019}. 

In Fig.~\ref{fig:ExpPopEv}, we show the temporal evolution of the relative populations $F_{1,1}$ and $F_{3,1}$ during $\mathrm{1.25~ms}$ in steps of $20~\mu\mathrm{s}$. Here, the driving strength after a linear increase during $0.5\,$ms is kept constant for $0.75\,$ms. For weak driving strength $f_{d} = 0.17$, shown in Fig.~\ref{fig:ExpPopEv}(a), we observe a moderate decrease of $F_{1,1}$ over the full modulation period, consistent with the suppression of the DW$_{1,1}$ phase observed in Ref.~ \cite{Georges2018}, while only a very small fraction of the atoms are transferred into $F_{3,1}$. For stronger driving, depicted in Fig.~\ref{fig:ExpPopEv}(b) and Fig.~\ref{fig:ExpPopEv}(c), with $f_{d} = 0.52$ and $f_{d} = 0.71$, respectively, the picture changes. Guided by the theoretically expected oscillatory dynamics discussed below, we interpret the observations of the occupations $F_{3,1}$ and $F_{1,1}$ as oscillations strongly damped by excessive heating. Fits with damped harmonic oscillations (black solid lines) let us extract the oscillation frequencies $\omega_{<}$ for $F_{3,1}$ in panels (a-c). Only the data points after the modulation strength reaches its maximal value at $0.5\,$ms are used. The frequencies thus determined are used as fixed parameters in the corresponding fits of $F_{1,1}$. We find $\omega_{<} \approx 0.8\,$kHz in (a), $\omega_{<} \approx 0.9\,$kHz in (b), and $\omega_{<} \approx 1.2\,$kHz in (c) with fairly large errors on the order of $20\,\%$ in (a) and $10\,\%$ in (b) and (c). Note the increase of $\omega_{<}$ with increasing $f_{d}$.

The interpretation of our experimental findings requires a few theoretical preparations. The equations of motion for the matter field $\Psi(y,z,t)$ and the cavity field ${\alpha}$ are given by Eqs. (57a) and (57b) in Ref.~\cite{Ritsch2013}:
\begin{align}\label{eq:eom}
i\hbar\frac{\partial {\Psi}(y,z,t)}{\partial t} &= \left(-\frac{\hbar^2}{2m}\nabla^2 + U_{\mathrm{dip}}(y,z,t)\right) {\Psi}(y,z,t) \\ \nonumber
i\frac{\partial {\alpha}}{\partial t} &= \left(-\delta_c + U_0\mathcal{B} -i\kappa\right){\alpha} + \sqrt{U_0\epsilon_{p}(t)}\,\Theta_{1,1} + i\xi.
\end{align}
Here, $\delta_c$ is the detuning between the pump frequency and the empty cavity resonance, $\mathcal{B}=\langle \mathrm{cos}^2(kz) \rangle$ is the bunching parameter and $\Theta_{1,1}=\langle \mathrm{cos}(kz)\mathrm{cos}(ky) \rangle$ is the DW$_{1,1}$ order parameter. The latter is an example of the general time-dependent order parameter
\begin{equation}
\label{eq:orderparam}
\Theta_{\mu,\nu}(t)=\int dydz~|{\Psi}(y,z,t)|^2 \mathrm{cos}(\mu ky)  \mathrm{cos}(\nu kz) \, .
\end{equation}
In Eq.~\eqref{eq:eom}, we neglect the effects of collisional atom-atom interactions. The time-dependent dipole potential $U_{\mathrm{dip}}$ due to the cavity and pump fields with mode functions $f(z)=\mathrm{cos}(kz)$ and $g(y)=\mathrm{cos}(ky)$, respectively, is 
\begin{align}
U_{\mathrm{dip}}(y,z,t)/\hbar= & U_0f(z)^2|\alpha|^2 + \epsilon_{p}(t) |g(y)|^2  \\ \nonumber
& + \sqrt{U_0\,\epsilon_{p}(t)}f(z)g(y)\left(\alpha+\alpha^{*} \right).
\end{align} 
The fluctuations of the cavity field are captured by the stochastic noise term $\xi(t)$ satisfying $\langle \xi^*(t)\xi(t') \rangle = \kappa\, \delta(t-t')$ \cite{Ritsch2013}. We employ the truncated Wigner approximation (TWA) \cite{Blakie2008,Polkovnikov2010,Carusotto2013} to simulate the dynamics of the system. By including initial quantum and stochastic noises, TWA can test the stability of nonequilibrium phases against inherent perturbations \cite{Cosme2018,Kessler2019}. For our numerical simulations, we use the same parameters as in the experiment. The equations of motion in Eq.~\eqref{eq:eom} are numerically solved by expanding the BEC wavefunction in the plane-wave basis ${\Psi}(y,z) = \sum_{\mu,\nu}{\phi}_{\mu,\nu}\,\mathrm{e}^{i \mu k y}\mathrm{e}^{i \nu k z}$, where ${\phi}_{\mu,\nu}$ is a single-particle momentum mode with occupation $n_{\mu,\nu}\equiv\langle \phi^{\dagger}_{\mu,\nu}\phi_{\mu,\nu} \rangle$.

\begin{figure}[!htbp]
\centering
\includegraphics[width=1\columnwidth]{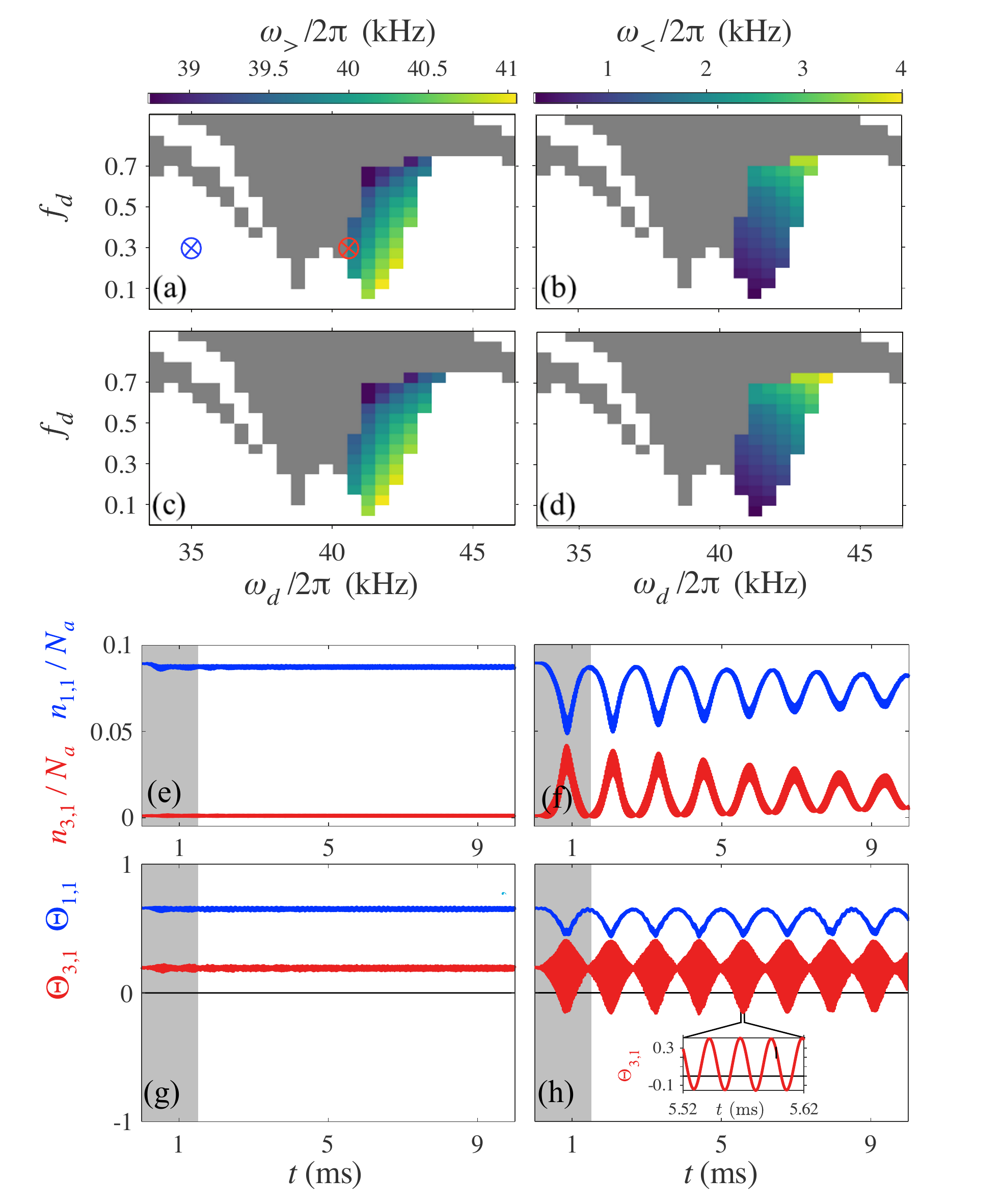}
\caption{The mean-field dynamics of $n_{1,1}$ and $n_{3,1}$ is anaylzed in (a) and (b). Gray and white regions show chaotic dynamics and prevailing DW$_{1,1}$ order, respectively. In the colored region persistent oscillations at the two frequencies $\omega_{>}$ and $\omega_{<}$ are found. The color scales show $\omega_{>}$ in (a) and $\omega_{<}$ in (b). In (c) and (d) analogous analysis is performed for the order parameters $\Theta_{1,1}(t)$ and $\Theta_{3,1}(t)$. TWA calculations are shown for $n_{1,1}/N_a$ and $n_{3,1}/N_a$ for off-resonant modulation in (e) and near-resonant modulation in (f). Corresponding order parameters $\Theta_{1,1}(t)$ and $\Theta_{3,1}(t)$ according to Eq.~\eqref{eq:orderparam} are shown in (g) and (h). The gray areas indicate the experimentally accessible time windows for driving before excessive heating sets in. The inset in (h) zooms into the short-time dynamics of $\Theta_{3,1}$. The modulation strength is fixed at $f_{d}=0.3$ and $\omega_{d}/2\pi=35.0~\mathrm{kHz}$ in (e, g) and $\omega_{d}/2\pi=40.5~\mathrm{kHz}$ in (f, h). The plots in (e, g) and (f, h) correspond to positions indicated by the blue and red circles in (a), respectively.}
\label{fig:ThDyn} 
\end{figure} 

We may now discuss the theoretically expected dynamics and compare it to the experimental observations. In each panel (a-d) of Fig.~\ref{fig:ThDyn}, we show the same rectangular region of the ($f_{d}$, $\omega_{d}$)-plane. In panels (a) and (b) three different regions can be recognized. The white region shows those locations, where $n_{3,1}$ is practically zero, while $n_{1,1}$ oscillates around some finite constant offset value with a small amplitude at the driving frequency $\omega_{d}$. This indicates a dominant DW$_{1,1}$ phase, slightly affected by non-resonant driving, as in the blue shaded region of Fig.~\ref{fig:ExpPhDg}(a). The gray region exhibits chaotic irregular dynamics of $n_{1,1}$ and $n_{3,1}$ at multiple frequencies. Finally, the colored islands in (a) and (b) show locations, where both $n_{1,1}$ and $n_{3,1}$ perform persistent regular oscillations, each at the same pair of distinct frequencies $\omega_{>}$ and $\omega_{<}$. The former ($\omega_{>}$) varies between 39~kHz and 41~kHz, shown in (a) via the color scheme at the upper edge of that panel, and the latter ($\omega_{<}$) between zero and 4~kHz shown in (b), correspondingly. As a consequence, in this region, the momentum spectrum of the atomic sample oscillates between the two cases illustrated in Fig.~\ref{fig:intro} (b) and (c). The value of $\omega_{>}$ is a function of system inherent parameters but also on $f_{d}$ and $\omega_{d}$, while $\omega_{<}$ satisfies the relation $\omega_{<} = \omega_{d} - \omega_{>}$. In Fig.~\ref{fig:ThDyn}(c) and (d), the same analysis as in (a) and (b) is performed, however, considering the order parameters $\Theta_{1,1}$  and $\Theta_{3,1}$. The same oscillation frequencies are found in the colored regions as in (a) and (b). This leads us to identify the dynamical DW order discussed here as DW$_{3,1}$. The colored island, signaling persistent dual-frequency dynamics, qualitatively corresponds to the red colored resonance lobe in the experimental phase diagram in Fig.~\ref{fig:ExpPhDg}. Note that, according to our calculations, the regime of persistent stable oscillations requires recoil resolution ($\kappa < 2 \,\omega_{\textrm{rec}}$) of the cavity (cf. Appendix D).

In order to account for realistic initial conditions and fluctuations as imposed by quantum noise, we have performed TWA simulations at two points in Figs.~\ref{fig:ThDyn}(a-d), one in the white region and one in the colored island, indicated by blue and red circles, respectively. In Figs.~\ref{fig:ThDyn}(e) and (f), TWA results for $n_{1,1}$ (blue line) and $n_{3,1}$ (red line) are plotted versus time for off-resonant (blue circle in (a)) and near-resonant (red circle in (a)) driving, respectively. In the off-resonant case, $n_{3,1}$ is practically zero while $n_{1,1}$ has a constant value close to 0.09. Zooming into this graph, one finds an oscillation with a tiny amplitude at the driving frequency $\omega_{d}$. This constitutes the case of light-induced coherence for intermediate modulation strength \cite{Cosme2018, Georges2018}. More interestingly, for near-resonant driving, the system develops periodic structures in space and time. Both populations $n_{1,1}$ and $n_{3,1}$ oscillate inversely at the frequency $\omega_{<} = \omega_{d} - \omega_{>}$ along with a much smaller oscillation at the frequency $\omega_{>}$. In the experiment, the corresponding oscillations at $\omega_{<}$ in Fig.~\ref{fig:ExpPopEv}(a-c) are strongly damped by heating and by dephasing and atom loss due to contact interaction, such that at most a single oscillation cycle is visible. The observed frequencies around $\omega_{<} \approx 1$~kHz together with the applied driving frequency of 40~kHz yield $\omega_{>} \approx 39\,$kHz. The fast oscillation at frequency $\omega_{>}$ is more clearly seen by analyzing the order parameter $\Theta_{3,1}$ associated with DW$_{3,1}$ order. In Fig.~\ref{fig:ThDyn} $\Theta_{3,1}$ is plotted for off-resonant (g) and near-resonant (h) modulation, respectively. The near-resonant case clearly shows persistent oscillatory dynamics at the frequencies $\omega_{<}$ and $\omega_{>}$. One may be tempted to interpret the DW$_{3,1}$ phase in terms of collective Rabi-dynamics at the frequency $\omega_{<}$ induced by the carrier and the sidebands of the pump beam via Raman two-photon coupling between the light-shifted momentum states $\{\pm 3, \pm 1\} \hbar k$ and the DW$_{1,1}$ state, assuming these are stationary and separated by the energy $\hbar \omega_{>}$. However, this picture falls short, since $\omega_{>}$ is itself a function of the driving parameters and the Raman coupled states exhibit self-organized polaritonic nature (cf. Appendix A). 

\textbf{Acknowledgments} This work was supported by the Deutsche Forschungsgemeinschaft (DFG) through project SFB-925 C5 and the Cluster of Excellence \textit{Advanced Imaging of Matter} (EXC 2056), Project No. 390715994.

\appendix

\begin{figure*}[!hbtp]
\centering
\includegraphics[width=2.0\columnwidth]{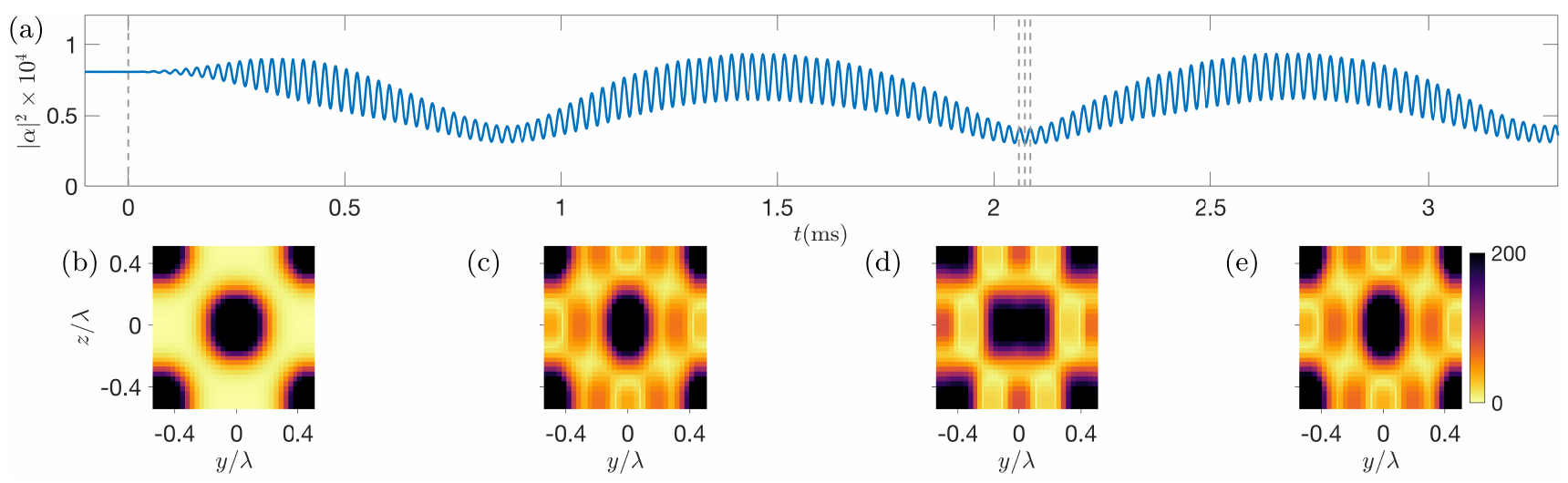}
\caption{Mean-field theoretical results for the same parameters as in Figs.~\ref{fig:ThDyn}(f) and \ref{fig:ThDyn}(h). (a) Dynamics of the cavity mode occupation. (b)-(e) Snapshots of the single-particle density (SPD) profile taken at instants of time denoted by the vertical gray dashed lines in (a) from left to right. The last three SPD profiles in (c)-(e) are taken over one modulation period.}
\label{fig:spd} 
\end{figure*} 

\section{Hamiltonian and Excitation of DW$_{3,1}$}

The Hamiltonian describing the atom-cavity system in the momentum basis is given by \cite{Cosme2018,Cosme2019}
\begin{align}\label{eq:hamilt}
	\hat{H}&/\hbar= -\delta_{\mathrm{C}}\hat{\alpha}^{\dagger}\hat{\alpha} + \frac{U_0}{4}\hat{\alpha}^{\dagger}\hat{\alpha}\sum_{n,m}\left( \hat{\phi}^{\dagger}_{n,m+2}\hat{\phi}_{n,m} + \mathrm{H.c.} \right) \\ \nonumber
	 &+\frac{U_0}{2}\hat{\alpha}^{\dagger}\hat{\alpha}\sum_{n,m}\hat{\phi}^{\dagger}_{n,m}\hat{\phi}_{n,m} + \omega_{\mathrm{rec}}\sum_{n,m}(n^2+m^2)\hat{\phi}^{\dagger}_{n,m}\hat{\phi}_{n,m}  \\ \nonumber
	 &- \frac{\omega_{\mathrm{rec}}}{4}\epsilon_p\sum_{n,m}\left( \hat{\phi}^{\dagger}_{n+2,m}\hat{\phi}_{n,m} + \mathrm{H.c.}\right)  \\ \nonumber
	 &-\frac{\omega_{\mathrm{rec}}}{2}\epsilon_p\sum_{n,m}\hat{\phi}^{\dagger}_{n,m}\hat{\phi}_{n,m} +\frac{\sqrt{\omega_{\mathrm{rec}} |U_0| \epsilon_p}}{4}\,(\hat{\alpha}^{\dagger}+\hat{\alpha}) \\ \nonumber
	 &\qquad \times \sum_{n,m}\left( \hat{\phi}^{\dagger}_{n,m}(\hat{\phi}_{n+1,m+1}+\hat{\phi}_{n+1,m-1}) +  \mathrm{H.c.} \right).
\end{align}
In the semiclassical limit of large $N_a$, the momentum and cavity modes can be treated as $c$ numbers leading to the following equations of motion derived within the truncated Wigner formalism:
\begin{align}\label{eq:eomsc}
i\frac{\partial {\phi}_{n,m}}{\partial t} &= \frac{\partial {H}}{\partial {\phi}^{*}_{n,m}}\\ \nonumber
i\frac{\partial {\alpha}}{\partial t} &= \frac{\partial {H}}{\partial {\alpha}^{*}}-i\kappa{\alpha} + i\xi.
\end{align}
In the DW$_{1,1}$ and DW$_{3,1}$ orders, the relevant modes are  $\phi_{\pm 1, \pm 1}$ and $\phi_{\pm 3, \pm 1}$. For brevity and motivated by symmetry, we can simply focus on the time evolution of $\phi_{3, 1}$. Neglecting irrelevant momentum modes, the corresponding equation of motion is
\begin{align}\label{eq:31}
i &\frac{\partial \phi_{3,1}}{\partial t} = 
	\omega_{\mathrm{rec}}\left(10+\frac{U_0}{2 \omega_{\mathrm{rec}}}|\alpha|^2-\frac{1}{2}\epsilon_p\right)\phi_{3,1} \\ \nonumber
	&+\frac{U_0}{4}|\alpha|^2\, \phi_{3,-1}+\frac{\sqrt{\omega_{\mathrm{rec}}|U_0|\epsilon_p}}{2}\mathrm{Re}(\alpha) \phi_{2,0} -\frac{1}{4}\omega_{\mathrm{rec}}\epsilon_p\, \phi_{1,1} .
\end{align}

The last two terms in Eq.~\eqref{eq:31} suggest that a periodic evolution of $\phi_{1,1}$ and $\phi_{2,0}$ may act as an effective driving for $\phi_{3,1}$. This leads to the parametric excitation of $\phi_{3,1}$ for near-resonant driving with respect to the natural frequency of $\phi_{3,1}$. Furthermore, Eqs.~\eqref{eq:hamilt} and \eqref{eq:31} reveal the excitation paths for $\phi_{\pm 3, \pm 1}$: (i) atoms with momenta $\{\pm 1, \pm 1\}\hbar k$ are scattered to $\{\pm 3, \pm 1\}\hbar k$ by the pump lattice, (ii) atoms with momenta $\{\pm 2, 0\}\hbar k$ absorb a pump photon followed by an emission into the cavity, and (iii) atoms with momenta $\{\pm 2, 0\}\hbar k$  absorb a cavity photon followed by an emission into the pump field. 
\begin{figure}[!htbp]
\centering
\includegraphics[width=1.0\columnwidth]{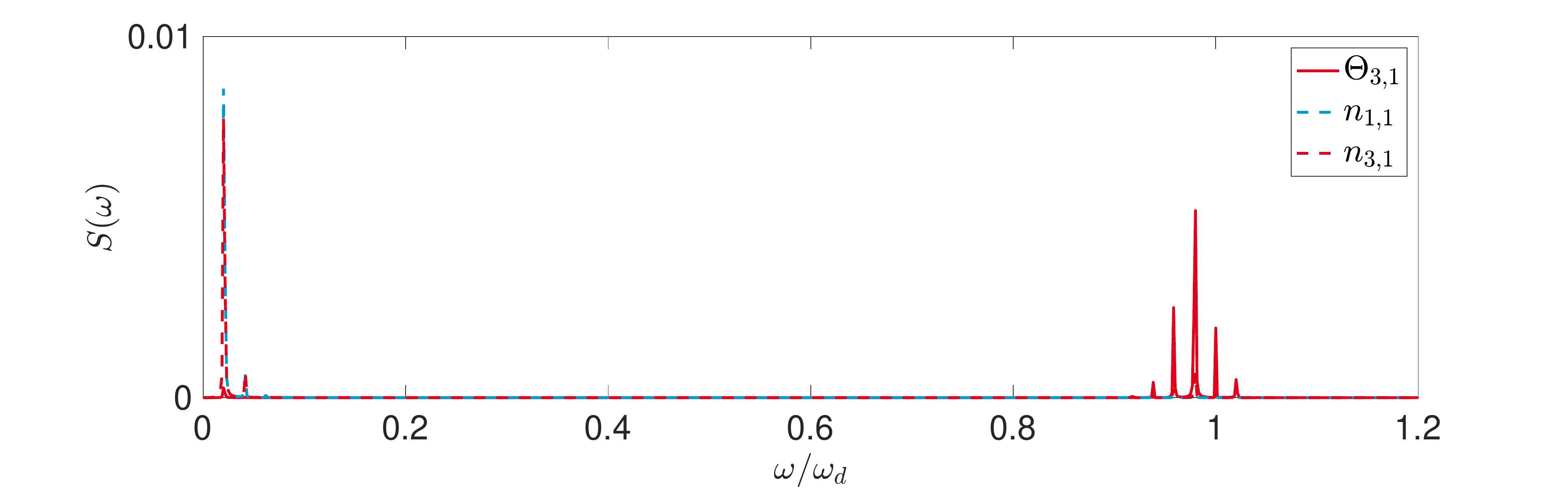}
\caption{Fourier spectra of the MF dynamics for $\Theta_{3,1}$, $n_{1,1}$, and $n_{3,1}$ corresponding to Figs.~\ref{fig:ThDyn}(f) and \ref{fig:ThDyn}(h)}
\label{fig:FFT} 
\end{figure} 
Since the number of atoms is conserved, at least in theory, the transfer of atoms from $\phi_{1,1}$ and $\phi_{2,0}$ to $\phi_{3,1}$ effectively decreases the strength of the parametric driving for $\phi_{3,1}$. This then leads to a depopulation of $\phi_{3,1}$ if the strength of the effective drive induced by oscillations in $\phi_{1,1}$ and $\phi_{2,0}$ falls below the critical value for the parametric resonance. As $\phi_{1,1}$ and $\phi_{2,0}$ recover atoms, the effective driving strength experienced by $\phi_{3,1}$ increases again and the cycle begins anew leading to coherent oscillations of $n_{3,1}$, such as those in Figs.~\ref{fig:ThDyn}(f) and \ref{fig:ThDyn}(h). The process described above highlights the importance of initializing in a DW$_{1,1}$ order since the excitation paths for $\phi_{3,1}$ rely on the occupation of the cavity mode $\alpha$ and $\phi_{\pm 1,\pm 1}$. In the experiment, atom loss arising from driving-induced heating interferes with the repopulation of $\phi_{1,1}$ and $\phi_{2,0}$, which is a key ingredient in the parametric excitation of $\phi_{3,1}$. This can explain the disappearance of $n_{1,1}$ and $n_{3,1}$ after 1 ms in Fig.~\ref{fig:ExpPopEv}.

In Fig.~\ref{fig:FFT}, we present an example of the Fourier spectra for $\Theta_{3,1}$, $n_{1,1}$, and $n_{3,1}$ in the DW$_{3,1}$ order. It shows the fast frequency $\omega_>$, which is the main frequency peak slightly red-detuned from the driving frequency, and the slow frequency $\omega_<$, which is the dominant peak close to $\omega=0$.

\begin{figure}[!htbp]
\centering
\includegraphics[width=1.0\columnwidth]{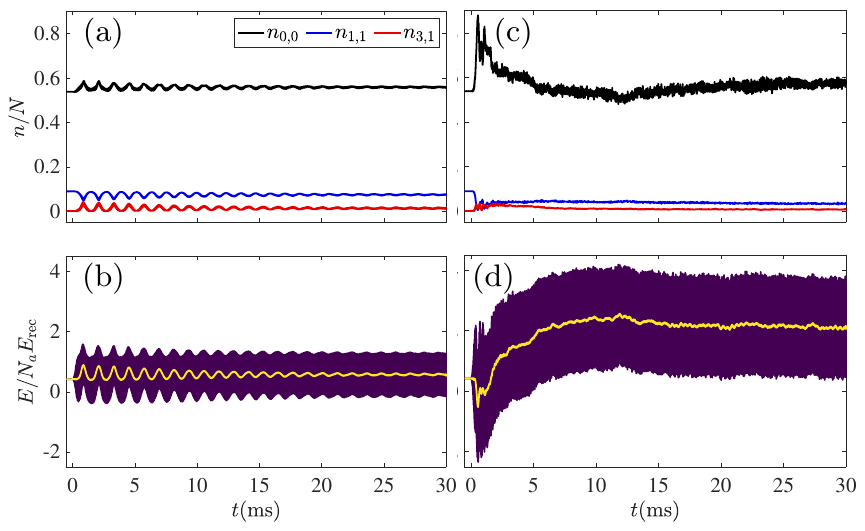}
\caption{TWA results of dynamics for (a),(b) $f_0=0.3$ and (c),(d) $f_0=0.9$ with fixed $\omega_d/2\pi=40~\mathrm{kHz}$. (a),(c) Dynamics of the occupation of $n_{0,0}$, $n_{1,1}$, and $n_{3,1}$. (b),(d) Corresponding time evolution of the expectation value of the Hamiltonian in Eq.~\eqref{eq:hamilt}. The dark line denotes the actual dynamics while the yellow line denotes the running average.}
\label{fig:energy} 
\end{figure} 

\section{Mean-field single-particle density profile}

To visualise the dynamical density wave order, we calculate the mean-field single-particle density (SPD) profile of the atomic ensemble,
\begin{align}
\rho(y,z) &\equiv \langle \Psi^{\dagger}(y,z)\Psi(y,z) \rangle \\ \nonumber
&= \sum_{n,m,n',m'} \phi^{*}_{n,m}\phi_{n',m'}e^{i(n-n')ky}e^{i(m-m')kz}.
\end{align}
Exemplary SPD profiles and dynamics of the cavity occupation for a DW$_{3,1}$ order are shown in Fig.~\ref{fig:spd}. Since the excitation of $\phi_{\pm 3,\pm 1}$ depends on the initial occupation of $\phi_{\pm 1,\pm 1}$, the $\mathbb{Z}_2$-symmetry breaking in the DW$_{1,1}$ order carries over in the DW$_{3,1}$ order. This explains the offset in the mean value of $\Theta_{3,1}$ in Fig.~\ref{fig:ThDyn}(h). 
The additional density modulations due to the occupation of $\phi_{\pm 3,\pm 1}$ are visible in Figs.~\ref{fig:spd}(c)-\ref{fig:spd}(e), which can be compared to the SPD of a typical DW$_{1,1}$ as shown in Fig.~\ref{fig:spd}(b). The oscillation of $\Theta_{3,1}$ between positive and negative values depicted in Fig.~\ref{fig:ThDyn}(h) manifest in the dynamical switching of the higher-order density grating (smaller period) from Fig.~\ref{fig:spd}(c) to Fig.~\ref{fig:spd}(e) over half a modulation cycle.

\begin{figure}[!htbp]
\centering
\includegraphics[width=1.0\columnwidth]{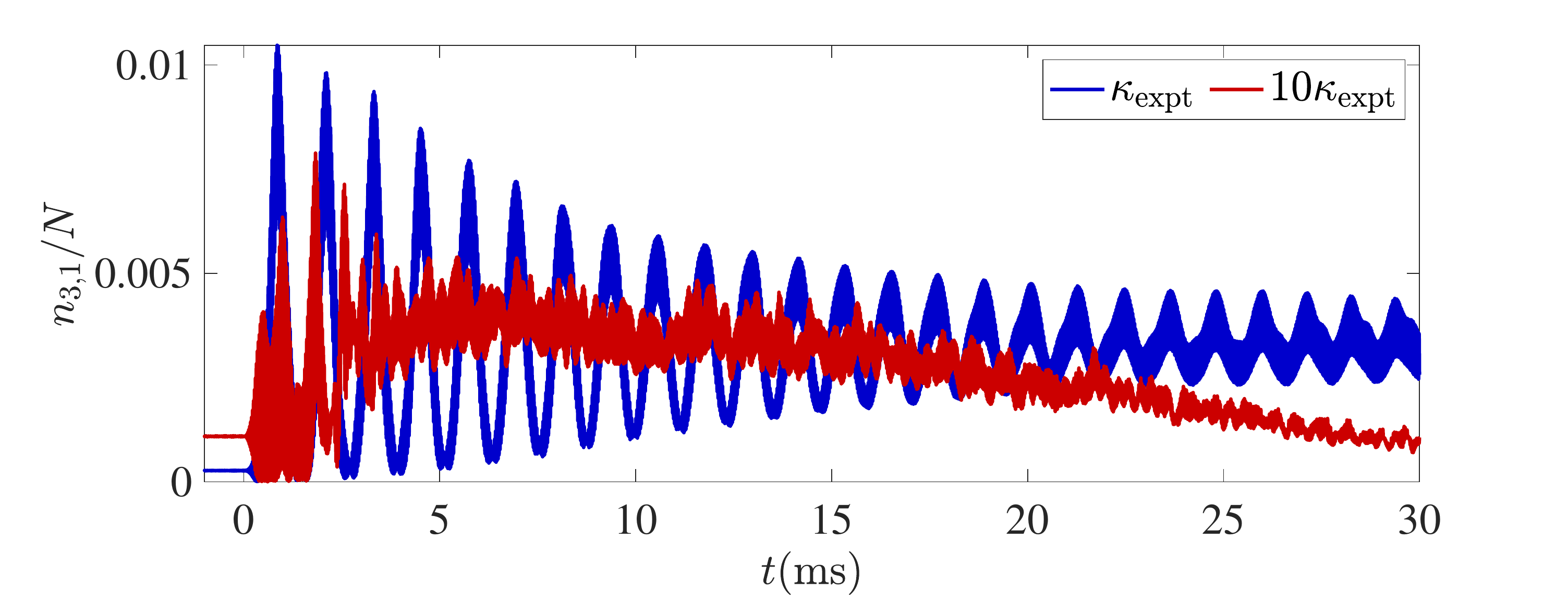}
\caption{Dynamics of $n_{1,1}$ from TWA for different cavity decay rate $\kappa$ with fixed $\delta_{\mathrm{eff}} = \omega_p - \omega_{c,\mathrm{eff}} = -2\pi \times 30~\mathrm{kHz}$. The pump strength $\epsilon$ is chosen such that $n_{1,1}$ is the same prior to modulation for all $\kappa$.}
\label{fig:kappa} 
\end{figure} 

\section{Heating}
We identify the gray areas in Fig.~\ref{fig:ThDyn}(a)-\ref{fig:ThDyn}(d) as regions dominated by chaotic dynamics, which in the experiment will contribute to heating and atom loss. In Fig.~\ref{fig:energy}, we present the dynamics of the relevant momentum modes and the expectation value of the Hamiltonian in Eq.~\eqref{eq:hamilt}, $\langle \hat{H} \rangle=E$, which can be used to indicate heating in the system. Note that while our numerical simulations show a small amount of recondensation, we did not observe such an effect in our experiment, where the coherent fraction of atoms is depleted by collisional decoherence and heating due to mechanical instability of the cavity and driving. For the DW$_{3,1}$ order, the energy per particle shown in Fig.~\ref{fig:energy} oscillates around its value prior to the modulation. On the other hand in the chaotic regime, aside from individual mean-field trajectories exhibiting irregular dynamics, the amplitude of oscillations in the energy is larger and its time average grows to almost 20 times its initial value (see Fig.~\ref{fig:energy}(d)). \\

\section{Significance of Recoil Resolution}
In the present work, the cavity operates in the recoil resolved regime where $\kappa < 2\omega_{\mathrm{rec}}$. The opposite regime $\kappa \gg 2\omega_{\mathrm{rec}}$ is employed in most experimental projects. In this regime, the cavity mode can be adiabatically eliminated as it simply follows the dynamics of the atoms. While a more detailed investigation on how the dynamical phase diagram (see Fig.~\ref{fig:ThDyn}) changes with increasing $\kappa$ is desirable, we briefly show here results for $\kappa = 10\, \kappa_{0} \gg 2\omega_{\mathrm{rec}}$, where $\kappa_{0}$ denotes the value of $\kappa$ used in this work. In Fig.~\ref{fig:kappa}, we compare the dynamics of $n_{3,1}$ for $\kappa = \kappa_{0}$ and $\kappa = 10\, \kappa_{0}$. For $10\, \kappa_{0}$, in addition to initial short-lived irregular dynamics, the occupation of $\phi_{\pm 3, \pm 1}$ decays, which is in contrast to the long-lived oscillatory behavior found for $\kappa = \kappa_{0}$. Our findings emphasize the importance of recoil resolution for long-lived oscillatory dynamics and persistent stability of the DW$_{3,1}$ order.

\end{document}